# Stabilization of higher-order vortices and multi-hump solitons in media with synthetic nonlocal nonlinearities

Yaroslav V. Kartashov,[1] Victor A. Vysloukh,[2] and Lluis Torner[1]

[1]ICFO-Institut de Ciencies Fotoniques, and Universitat Politecnica de Catalunya, Mediterranean Technology Park, 08860 Castelldefels (Barcelona), Spain

[2]Departamento de Fisica y Matematicas, Universidad de las Americas – Puebla, Santa Catarina Martir, 72820, Puebla, Mexico

We address the evolution of higher-order excited states, such as vortex and multi-hump solitons, in nonlocal media with synthetic, competing focusing and defocusing nonlinearities with different nonlocal transverse scales. We reveal that introduction of suitable competing effects makes possible the stabilization of vortex solitons with topological charge $m > 2$, as well as one-dimensional multi-hump solitons with number of humps $p > 4$, all of which are highly unstable in natural nonlocal materials with focusing nonlinearities.

Nonlocality of the nonlinear response is a common physical phenomenon that is encountered in a variety of materials [1]. Strongly nonlocal nonlinear responses may occur, for example, in liquid crystals featuring a long-range reorientational nonlinearity [2-4], in atomic vapors where atomic diffusion causes transport of excitation away from the light-matter interaction region [5], in thermal materials where heat diffusion results in boundary-dependent redistribution of refractive index [6-8], in condensates of dipolar particles oriented by an external field and interacting via long-range anisotropic dipole-dipole interactions [9], in plasmas [10] and in photorefractive materials [11]. In most cases nonlocality appears as a result of long-range interactions or diffusion-type processes. Since in such media excitation in one spatial point affects the material properties in a nonlocal way, the propagation of light depends strongly on the beam width and on the particular shape of the response function. In many cases the photoinduced refractive index profile extends far beyond the illuminated region, resulting in long-rate transverse effects as well as in the stabilization of nonlinear light patterns that are self-destroy in local media [1]. Thus, nonlocality may result in suppression of azimuthal instabilities and stabilization of ring-profile vortex solitons [12-19], it can change the sign of interaction between out-of-phase spots, or make possible the formation of multi-hump solitons [20-30]. Vortex solitons with the lowest topological charges have been observed in thermal media [6], while the simplest multi-hump solitons have been observed in liquid crystals [22] and in thermal media [27]. Nevertheless, a general property of nonlocal nonlinearities explored so far experimentally in natural materials, is that they can not support the propagation of stable higher-order excited states such as vortices with high topological charges and multi-hump solitons with large number of humps. Only the vortices with charges $m \leq 2$ have been found to be stable in liquid-crystal and thermal media [15,18], while stable one-dimensional solitons in such materials can only exhibit $p \leq 4$ humps [23].

In this work we reveal that properly synthesized metamaterials with competing nonlocal nonlinearities exhibiting both, focusing and defocusing contributions with different nonlocality degrees, do allow stabilization of higher-order solitons. We conjecture that such synthetic nonlinearities may be achievable in composites made of materials with different physical properties, such as, e.g. thermal materials with different signs of thermo-optic coefficients, or in materials with two simultaneous physically different mechanisms of nonlocality, such as dye-doped liquid crystals. In particular, the existence of fundamental solitons in liquid crystals with competing focusing reorientational and defocusing thermal contributions to total nonlinear refractive index was recently demonstrated experimentally in [31]. Our salient discovery is that addition of a suitable defocusing contribution to the effective nonlocal response affords remarkably new phenomena, such as the stabilization of vortices with topological charges $m > 2$ and one-dimensional solitons with more than $p = 4$ humps.

To model the beam propagation in a medium with competing focusing and defocusing nonlocal nonlinearities we utilize the following system of equations for the dimensionless amplitude of light field $q$ and contributions to the refractive index $n_1$ and $n_2$:

$$\begin{aligned}
i\frac{\partial q}{\partial \xi} &= -\frac{1}{2}\Delta_\perp q - q(n_1 + n_2), \\
n_1 - d_1 \Delta_\perp n_1 &= \sigma_1 |q|^2, \\
n_2 - d_2 \Delta_\perp n_2 &= \sigma_2 |q|^2.
\end{aligned} \qquad (1)$$

Here $\Delta_\perp = \partial^2/\partial r^2 + (1/r)\partial/\partial r + (1/r^2)\partial^2/\partial\varphi^2$ is the transverse Laplacian, $r$ and $\varphi$ are the radius normalized to characteristic transverse scale and azimuthal angle, respectively, $\xi$ is the longitudinal coordinate normalized to diffraction length, parameters $\sigma_{1,2}$ characterize the strength and sign of nonlinearity components (positive values correspond to focusing, while negative values correspond to defocusing), while $d_{1,2}$ stand for degrees of nonlocality of nonlinear response. At $d_{1,2} \to 0$ one recovers a cubic nonlinear medium, while opposite limit $d_{1,2} \to +\infty$ corresponds to strongly nonlocal response.

First we consider vortex soliton solutions of Eq. (1) that can be presented in the form $q = w(r)\exp(ib\xi + im\varphi)$, $n_{1,2} = n_{1,2}(r)$, where $b$ is the propagation constant and $m$ is the topological charge. Notice that the total refractive index writes as

$$n(r) = n_1(r) + n_2(r) = 2\pi \int_0^\infty [\sigma_1 G_1^0(r,\rho) + \sigma_2 G_2^0(r,\rho)]|q(\rho)|^2 \rho d\rho,$$

$$G_{1,2}^k(r,\rho) = \frac{1}{2\pi d_{1,2}} K_k(d_{1,2}^{-1/2}\rho) I_k(d_{1,2}^{-1/2}r) \ \text{ for } \ r \leq \rho, \qquad (2)$$

$$G_{1,2}^k(r,\rho) = \frac{1}{2\pi d_{1,2}} K_k(d_{1,2}^{-1/2}r) I_k(d_{1,2}^{-1/2}\rho) \ \text{ for } \ r > \rho,$$

where $I_k$ and $K_k$ are modified Bessel functions of first and second kind. Notice that for $\text{sgn}(\sigma_1) \neq \text{sgn}(\sigma_2)$ the response function $\sigma_1 G_1^0 + \sigma_2 G_2^0$ may exhibit a complex shape and change its sign, in contrast to the usual monotonic focusing response. Because the shape of the response function is a crucial factor that determines the stability of vortices and multi-hump solitons our intuition is that modifications in response function may substantially affect soliton stability. At $\sigma_2 = 0$ the system (1) describes the response of liquid crystals [2-4] and plasmas [10]. Further we fix $\sigma_1 = 1$ and vary $d_{1,2}$ and $\sigma_2 < 0$.

We are interested in the stabilization of vortex solitons that are unstable when $\sigma_2 = 0$ and thus we look for perturbed solutions $q = [w + u\exp(ik\varphi) + v^*\exp(-ik\varphi)]\exp(ib\xi + im\varphi)$ of Eq. (1), where $u, v \sim \exp(\delta\xi)$ are small perturbations and $k$ is an azimuthal perturbation index. Linearization of Eq. (1) around the stationary solutions yields an eigenvalue problem from which the complex perturbation growth rate $\delta = \delta_r + i\delta_i$ might be found:

$$i\delta u = -\frac{1}{2}\left(\frac{d^2u}{dr^2} + \frac{1}{r}\frac{du}{dr} - \frac{(m+k)^2 u}{r^2}\right) - un + bu - w\delta n,$$
$$-i\delta v = -\frac{1}{2}\left(\frac{d^2v}{dr^2} + \frac{1}{r}\frac{dv}{dr} - \frac{(m-k)^2 v}{r^2}\right) - vn + bv - w\delta n. \qquad (3)$$

Here $\delta n = 2\pi \int_0^\infty (\sigma_1 G_1^k + \sigma_2 G_2^k) w(u+v)\rho d\rho$ is the refractive index perturbation.

First, we briefly describe the properties of vortex solitons for $\sigma_2 = 0$. For sufficiently high $d_1$ values the width of the refractive index profile substantially exceeds the one of solitons [Figs. 1(a) and 1(b)], see also [15]. The refractive index exhibits a plateau around $r = 0$ whose width is comparable with the vortex radius. Such plateau expands with growing $m$. The soliton power $U = 2\pi \int_0^\infty |q|^2 r dr$ grows monotonically with $b$ [Fig. 1(c)] and vanishes at $b \to 0$. At fixed $b$, vortex solitons with higher charges carry a higher power. Figure 2 shows the real part of the perturbation growth rate versus $b$ for different charges $m$ and azimuthal indices $k$. Vortices with charges 1 and 2 become stable when $b$ exceeds a critical value [Figs. 2(a) and 2(b)], which diminishes with increasing nonlocality degree $d_1$. Note that stability of $m = 2$ vortex is in contrast to result reported in [15]. Anyway, the point is that in this system all vortices with charges $m > 2$ are unstable. The most destructive perturbations correspond to the azimuthal indices $k = 3$ for $m = 3$ [Fig. 2(c)], and $k = 3, 4$ for $m = 4$ [Fig. 2(d)], etc. These results suggest that $m = 2$ is the maximal possible charge of stable vortices for liquid-crystal-like response functions and in thermal media (see Ref. [18]). Even azimuthal indices for the most modulationally destructive perturbations are alike in both cases. Notice that increasing the degree of nonlocality $d_1$ of the nonlinear response in our system results in a reduction of the corresponding perturbation growth rates for unstable vortices. Thus, when $d_1 \to \infty$, all instability growth rates vanish. However, at finite $d_1$, when the vortex soliton is unstable, the instability persists for all propagation constant values.

Next we introduce a synthetic defocusing component ($\sigma_2 = -0.2$) in the nonlinear response. One can see from Eq. (2) that at $d_2 \to \infty$ the response function $G_2^0$ becomes very broad and that the defocusing nonlinearity does not impact the total refractive index distribution [compare Fig. 3(a) obtained at $d_2 = 50$ with Fig. 1(b) corresponding to $\sigma_2 = 0$]. In contrast, when $d_2 \to 0$ the soliton shape is strongly affected by the defocusing nonlinearity even though $|\sigma_2| < |\sigma_1|$. Also, diminishing $d_2$ results in the monotonic increase of the soli-ton power and in a clearly visible stretching of the soliton shape [Fig. 3(b)]. The total refractive index acquires a local maximum at $r = 0$ in contrast to the case $\sigma_2 = 0$, while the $n(r)$ dependence folds notably around the soliton intensity maximum.

The central result of this work is that, as consequence of the above, addition of a small defocusing nonlinearity results in the *stabilization of vortices with high charges*. This is illustrated on Fig. 4(a) where the $\delta_r(b)$ dependence for the most destructive perturbation ($k = 3$) for $m = 3$ vortex is shown for different $d_2$ values. Decreasing $d_2$ changes the character of this dependence and results in the appearance of a stability domain at $b \geq b_{\text{cr}}$, when vortex amplitude becomes sufficiently high. Notice, that in the case of usual focusing nonlocal response the stabilizing effect of increased amplitude on lowest-order vortices in nonlocal medium was discussed analytically in [32]. Analogously, if one fixes $b$ and decreases $d_2$ the growth rates for all $k$ vanish when $d_2$ becomes less than a critical value $d_2^{\text{cr}}$. It should be stressed that in simulations we tested growth rates for azimuthal perturbation indices up to $k = 20$ which gives a strong numerical evidence of vortex stabilization because existence of destructive perturbations with $k > 20$ is unlikely. The vanishing of growth rates with decrease of $d_2$ takes place for $m = 3$ [Fig. 4(b)], $m = 4$ [Fig. 4(c)], and all vortices with higher charges that we investigated (here we limited ourselves to vortices with $m$ up to 10). The perturbation that disappear last with decreasing $d_2$ turns out to depend mostly on the value of $b$: For the parameters of Fig. 4 the most long-living perturbation is the one with $k = 2$. We found that the critical $d_2$ value is a monotonically increasing function of $b$ [Fig. 4(d)], but there exist a minimal propagation constant below which one can not achieve vortex stabilization even at $d_2 = 0$. Such value of $b$ increases monotonically with decreasing $d_1$. We found that the higher the charge of the vortex the larger the critical $b$ value, but the stability domain can always be found no matter how large is $m$. All these predictions of the linear stability analysis were confirmed by comprehensive direct simulations of the perturbed vortex solitons with high topological charges in Eq. (1). Notice that a hint about the stabilizing action of local small defocusing nonlinearity was encountered in [16] in a model for dipolar condensates for $m = 1$ vortices, which however can be stable in the absence of defocusing contributions too. It should be stressed that the mechanism of nonlocal stabilization (i.e., suppression of the amplification of azimuthal perturbations due to the fact that the refractive index is determined by the intensity in the surroundings of the point where perturbations occur) reported here is substantially different from mechanism resulting in stabilization of vortices in local cubic-quintic medium, where the nonlinear response changes its sign when the soliton amplitude becomes sufficiently large and the refractive index acquires a flat-top shape.

Importantly, stabilization of higher-order solitons in media with the complex response functions that we put forward here appears to be a more general phenomenon. Namely, we found that one can achieve such stabilization not only for vortices, but also for soliton trains, or multi-hump solitons sustained by the one-dimensional version of Eq. (1). When $\sigma_2 = 0$, all such solitons with more than $p = 4$ humps were found to be unstable [23], a result that holds in both liquid-crystal and thermal media. However, when $\sigma_2 < 0$, the response function $\sigma_1 G_1(\eta) + \sigma_2 G_2(\eta)$ [here $G_{1,2}(\eta) = (1/2d_{1,2}^{1/2})\exp(-|\eta|/d_{1,2}^{1/2})$ and $\eta$ is the transverse coordinate] may develop a dip in the center and can even change its sign with decreasing $d_2$. We found that such a feature causes the stabilization of multi-hump solitons with higher number of humps. An illustrative example of stable five-hump soliton is shown in Fig. 5(a). The stability domain for such solitons is given by $d_2^{\mathrm{low}} \leq d_2 \leq d_2^{\mathrm{upp}}$, i.e. addition of almost local defocusing nonlinearity may not result in stabilization of multi-hump soliton [Fig. 5(b)]. This is in contrast to vortices that can stabilized at very small $d_2$ values, as discussed above. The domain of stability for multi-hump solitons gradually turns out to shrink with decreasing $b$ and vanishes completely below a critical value. Here we limited ourselves to solitons with $p$ up to 10, and in all cases stabilization was found to be achievable.

Therefore, in summary, we revealed that higher-order excited states, such as vortex and multi-hump solitons, that are unstable in natural nonlocal media, can be made completely stable in properly synthesized nonlocal materials featuring both focusing and defocusing nonlocal contributions to the refractive index. The competing nonlocality can dramatically modify the stability of higher-order solitons, making possible that vortices with topological charges $m > 2$ and multi-hump solitons with number of humps $p > 4$ are stable.

# Figure captions

Figure 1. Profiles of vortex solitons with (a) $m = 1$ and (b) $m = 3$ at $b = 6$. (c) Power versus propagation constant for solitons with $m = 1$ and $3$. Points marked by circles in (c) correspond to solitons in Figs. 1(a) and 1(b). In all cases $d_1 = 15$ and $\sigma_2 = 0$. All quantities are plotted in arbitrary dimensionless units.

Figure 2. $\delta_r$ versus $b$ for vortices with (a) $m = 1$, (b) $m = 2$, (c) $m = 3$, and (d) $m = 4$. In all cases $d_1 = 15$ and $\sigma_2 = 0$. All quantities are plotted in arbitrary dimensionless units.

Figure 3. Profiles of vortex solitons corresponding to (a) $d_2 = 50$ and (b) $d_2 = 0.6$ at $b = 6$, $m = 3$, $d_1 = 15$, $\sigma_2 = -0.2$. All quantities are plotted in arbitrary dimensionless units.

Figure 4. (a) $\delta_r$ versus $b$ for perturbation with $k = 3$ at $m = 3$. $\delta_r$ versus $d_2$ for vortex solitons with (b) $m = 3$, $b = 5$ and (c) $m = 4$, $b = 9$. (d) Critical $d_2$ value for complete stabilization versus $b$. In all cases $d_1 = 5$ and $\sigma_2 = -0.2$. All quantities are plotted in arbitrary dimensionless units.

Figure 5. (a) Profile of stable five-hump soliton at $b = 12$, $d_2 = 0.4$. (b) Critical $d_2$ values for stabilization of five-hump solitons versus $b$. In all cases $d_1 = 5$ and $\sigma_2 = -0.2$. All quantities are plotted in arbitrary dimensionless units.

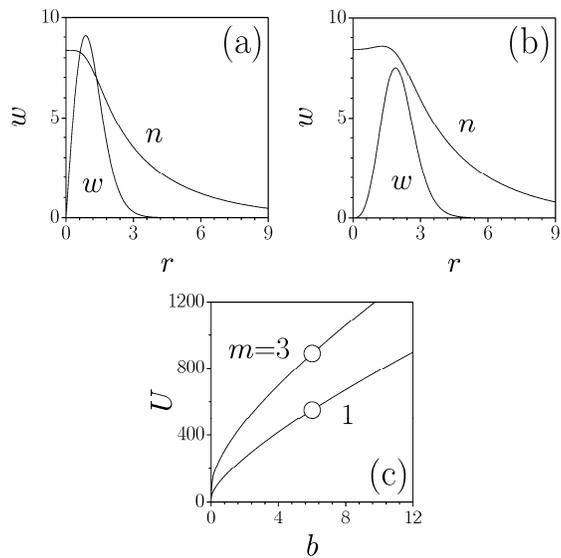

Figure 1.   Profiles of vortex solitons with (a) $m = 1$ and (b) $m = 3$ at $b = 6$. (c) Power versus propagation constant for solitons with $m = 1$ and $3$. Points marked by circles in (c) correspond to solitons in Figs. 1(a) and 1(b). In all cases $d_1 = 15$ and $\sigma_2 = 0$. All quantities are plotted in arbitrary dimensionless units.

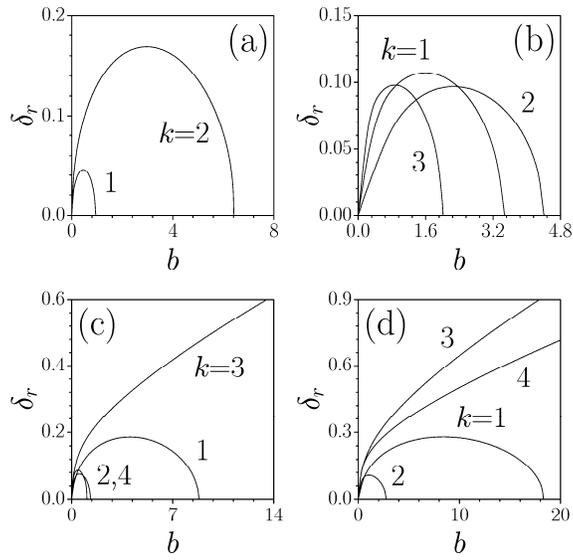

Figure 2. $\delta_r$ versus $b$ for vortices with (a) $m = 1$, (b) $m = 2$, (c) $m = 3$, and (d) $m = 4$. In all cases $d_1 = 15$ and $\sigma_2 = 0$. All quantities are plotted in arbitrary dimensionless units.

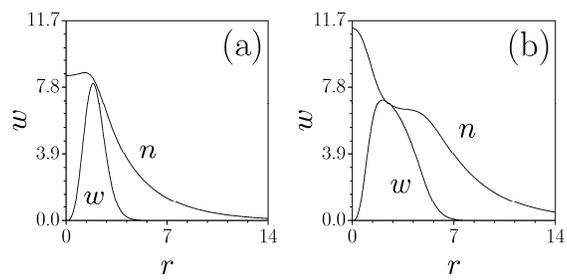

Figure 3. Profiles of vortex solitons corresponding to (a) $d_2 = 50$ and (b) $d_2 = 0.6$ at $b = 6$, $m = 3$, $d_1 = 15$, $\sigma_2 = -0.2$. All quantities are plotted in arbitrary dimensionless units.

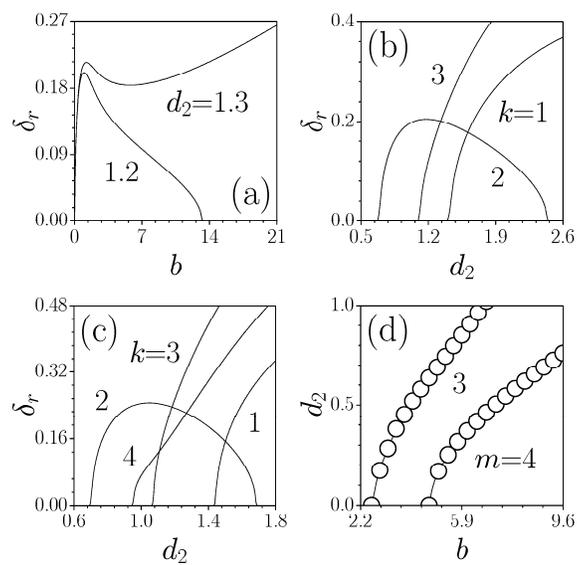

Figure 4. (a) $\delta_r$ versus $b$ for perturbation with $k = 3$ at $m = 3$. $\delta_r$ versus $d_2$ for vortex solitons with (b) $m = 3$, $b = 5$ and (c) $m = 4$, $b = 9$. (d) Critical $d_2$ value for complete stabilization versus $b$. In all cases $d_1 = 5$ and $\sigma_2 = -0.2$. All quantities are plotted in arbitrary dimensionless units.

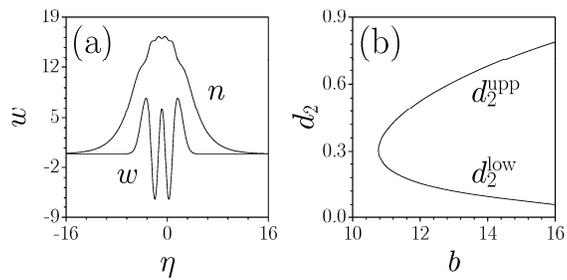

Figure 5.  (a) Profile of stable five-hump soliton at $b = 12$, $d_2 = 0.4$. (b) Critical $d_2$ values for stabilization of five-hump solitons versus $b$. In all cases $d_1 = 5$ and $\sigma_2 = -0.2$. All quantities are plotted in arbitrary dimensionless units.